\documentclass[usegraphicx,usenatbib]{mn2e}
\begin{document}

\title[ NGC 5548 ``Breathes'' ]
{ Photoionized H$\beta$ Emission in NGC 5548: It Breathes! }

\author[Cackett \& Horne]
{
Edward M. Cackett$^1$\thanks{emc14@st-andrews.ac.uk} and Keith Horne$^1$,
\\ $^1$School of Physics and Astronomy,
	University of St.~Andrews,
	KY16 9SS, Scotland, UK
}

\date{Accepted . Received ; 
in original form }

\maketitle

\begin{abstract} 

Emission-line regions in active galactic nuclei
and other photoionized nebulae should become larger in size
when the ionizing luminosity increases. 
This ``breathing'' effect is observed for the H$\beta$ emission in NGC~5548
by using  H$\beta$ and optical continuum lightcurves from the
13-year 1989-2001 AGN~Watch monitoring campaign.
To model the breathing, we use two methods to fit the 
observed lightcurves in detail: (i) parameterized models and, (ii) the
\texttt{MEMECHO} reverberation mapping code.
Our models assume that optical continuum variations track the
ionizing radiation, and that the H$\beta$ variations respond  
with time delays $\tau$ due to light travel time.
By fitting the data using a delay map $\Psi(\tau,F_c)$ that
is allowed to change with continuum flux $F_c$, 
we find that the strength of the H$\beta$ response 
decreases and the time delay increases with ionizing luminosity.
The parameterized breathing models allow the time delay and the H$\beta$ flux
to depend on the continuum flux so that, $\tau \propto
F_c^{\beta}$ and $F_{H\beta} \propto F_c^{\alpha}$.
Our fits give $0.1 < \beta < 0.46$ and $0.57 < \alpha < 0.66$.
$\alpha$ is consistent with previous work by \citet{gilbertp03} and
\citet*{goadkk04}.
Although we find $\beta$ to be flatter than previously determined by
\citet{peterson02} using cross-correlation methods, it is closer to the predicted values from recent theoretical work by \citet{koristagoad04}.

\end{abstract}

\begin{keywords}
galaxies: active -- galaxies: individual (NGC 5548) --
galaxies: nuclei -- galaxies: Seyfert
\end{keywords}

\section{Introduction}
\label{sec:intro}

Photoionization models predict the sizes of Str\"{o}mgren zones
for H~II regions and planetary nebulae
ionized by hot stars of various luminosities and spectral types.
Higher luminosity can maintain a larger mass of ionized gas.
Dynamical tests of photoionization models are rare.
The ionizing stars evolve in luminosity at rates too slow
for humans to directly observe changes in radius.
However, active galacic nuclei (AGN) vary on much shorter time-scales
(days to months).  Rapid variations in the ionizing luminosity emerging
from an AGN should cause the photoionized region to expand and contract.
This `breathing' of the emission-line region is an interesting
test of photoionization models.

Although the time-scales of the variations are convenient to human observers,
unfortunately, the angular sizes of the broad emission-line regions
are too small to be resolved directly.  A nebula 100 light days across at a
distance of 100 Mpc spans only 180 micro-arcseconds.
Fortunately, light travel times within the nebula
introduce time delays for any changes in the line emission.
The ionizing radiation from the innermost regions of an AGN is reprocessed
by gas in the surrounding broad line region (BLR).  As the central
source varies, spherical waves of heating and ionization, cooling and
recombination, expand at the speed
of light through the BLR.  A change in ionization causes a
corresponding change in the reprocessed emission.  In AGNs, the recombination
timescales of the gas in the BLR is very short compared to the light travel
time, so the delay seen by a distant observer is dominated by the light travel
time.  Hence, we see line
emission correlated with the continuum but with a time delay, $\tau$.  
A gas cloud 1 light day behind the ionizing source will
be seen to brighten 2 days after the ionizing source flux rises.
Thus, we can use light travel time delays to measure the size
of the region that is responding to variations in the ionizing flux, where
the reverberation radius is $\langle R\rangle \approx \langle\tau\rangle c$. 
Echo mapping, or reverberation mapping  \citep{blandmckee82}, aims to use this
correlated variability to determine the kinematics and structure
of the BLR, as well as the mass of the central supermassive black hole
\citep[e.g.][and references therein]{peterson93,petersonetal04}.

The nearby ($z = 0.017$) Seyfert 1 galaxy NGC~5548 has been intensively
monitored in the optical range for 13 years (1989-2001) by the international
\textit{AGN~Watch}\footnote{http://www.astronomy.ohio-state.edu/$\sim$agnwatch/}
consortium \citep[e.g.][]{peterson02}.  Those data spanning the source in a
wide range of luminosity states are ideal for searching
for this `breathing' effect.  In this paper, we investigate the luminosity
dependence of the H$\beta$ emission and present two methods of fitting
the data, accounting for the `breathing' using (i) parameterized models and
(ii) the reverberation mapping code \texttt{MEMECHO}
\citep*{hornewp91,horne94}.
Although previous work by \citet{petersonetal99,peterson02,gilbertp03,goadkk04}
has studied the luminosity dependence of the H$\beta$ emission in NGC~5548, this
study applies alternative techniques to characterise the `breathing'.
In $\S$\ref{sec:echomap} we describe the echo
mapping techinque and discuss the luminosity dependence of the emission-line
lightcurve. In $\S$\ref{sec:param} we present the luminosity dependent
parameterized models followed by the \texttt{MEMECHO} method in
$\S$\ref{sec:memecho}.  The results of these methods of fitting the data, and
their implications, are discussed in $\S$\ref{sec:discuss} and we summarize our
main findings in $\S$\ref{sec:conc}.

\section{Echo Mapping}\label{sec:echomap}

The emission line flux, $F_l(t)$, that we see at each time, $t$, is driven by
the continuum variations, $F_c(t)$, and arises from a range of time delays,
$\tau$.  The emission line lightcurve is therefore a delayed and blurred
version of the continuum lightcurve.
In the usual linearized echo model the line lightcurve is modelled as
\begin{equation}
  F_c(t) = \bar{F_c} + \Delta F_c(t)
\end{equation}
\begin{equation}
F_l(t) = \bar{F_l} + \int_0^{\tau_{\rm max}}
	\Psi\left( \tau \right) \Delta F_c(t-\tau) d\tau
\ .
\label{eq:model}
\end{equation}
where $\Psi(\tau)$ is the transfer function, or delay map.  We can adopt a
continuum background level $\bar{F_c}$, somewhat arbitrarily,
at the median of the observed continuum fluxes.
$\bar{F_l}$ is then a constant background line flux that would be produced
if the continuum level were constant at  $\bar{F_c}$.

A simple way of determining the size of the emission-line region is to
determine the time delay (or `lag') between the line and continuum lightcurves
using cross-correlation.  Taking the centroid of the cross-correlation function
(CCF) as the lag gives a luminosity-weighted radius for the BLR
\citep{robinsonperez90}.
However, the cross-correlation function is a convolution
of the delay map, $\Psi(\tau)$, with the auto-correlation function (ACF) of the
driving continuum lightcurve.
It is therefore possible that changes in the
measured cross-correlation lag arise from changes in the
continuum auto-correlation function rather than in the
delay map \citep*[e.g.][]{robinsonperez90,perezetal92a,welsh99}.
If the continuum variations become slower,
a sharp peak at low time-delay in the delay distribution will be
blurred by the broader auto-correlation function and the peak of the
cross-correlation function will be shifted to larger delays
\citep{netzermaoz90}.
A typical delay map may have a rapid rise to a peak at
small lag, and a long tail to large lags.
The asymmetric peak in $\Psi(\tau)$, 
will shift toward its longer wing when blurred by the
auto-correlation function.  Thus the lag measured by cross-correlation analysis
depends not only on the delay distribution, $\Psi\left(\tau\right)$, but also on
the characteristics (ACF) of the continuum variations.

Previous analysis of the \textit{AGN Watch} data for NGC 5548 by
\citet{peterson02} determined the H$\beta$ emission-line lag relative
to the optical continuum, on a year by year basis, using cross-correlation.
These authors find that the
lag increases with increasing mean continuum flux.
To improve upon the CCF analysis we use the
echo mapping technique to fit the lightcurves in detail.  However,
the linearized echo model (Eq.~\ref{eq:model}) is appropriate only when the
delay map is independent of time (static).  In this paper we extend the model to
search for changes in the delay map with luminosity.

\subsection{Luminosity-Dependent Delay Map}\label{sec:lumdep}

The above linearized echo model (Eq.~\ref{eq:model}) assumes that the line
emissivity and continuum flux can be related by a linear function and thus is
appropriate only for responses that are static.
In principle, the delay map may change with time, for example,
due to motion, or changes in quantity, of line-emitting gas within the system.
The delay map may also change with luminosity.  In the ``local optimally
emitting clouds'' (LOC) model \citep{baldwinetal95} at each time delay there is
a variety of gas clouds with differing properties, and those most efficient at
reprocessing tend to
dominate the line flux emerging from the region.  A change in ionizing
luminosity induces a change in the efficiency of reprocessing at each place in
the region and so the time delay at which the line emission is dominant will
change. 
When a cloud is partially ionized its response may initally be large so that
increasing luminosity increases the depth of the ionized zone on the face of
the cloud.  Once the cloud becomes completely ionized, however, further
increases in ionizing flux are less effectively reprocessed.
The line flux saturates, and may even decrease
with increasing ionizing flux due to either ionization or decline in the
recombination coefficients caused by an increase in gas temperature
\citep*{obriengg95}.

To account for these effects, we generalise the
echo model by
allowing the delay map to be luminosity-dependent,
$\Psi(\tau,F_c)$.
The response we see at time $t$ from a parcel of
emission-line gas located at time delay $\tau$ is set by 
the luminosity of the nucleus that we saw at the earlier time $t - \tau$.
Thus, \begin{equation}
F_l(t) = \bar{F_l} + \int_0^{\tau_{\rm max}}
        \Psi\left[ \tau, F_c(t-\tau) \right]
        \Delta F_c(t-\tau)
        d\tau
\ .
\label{eq:lumdep}
\end{equation}

\subsection{Luminosity dependence of H$\beta$ flux}\label{sec:hbdep}

A well-established correlation between continuum and emission-line properties is
the `Baldwin' effect \citep*{baldwin77,osmerpg94} where, in different AGN, broad
emission-line equivalent width is observed to decrease with increasing 
continuum level.  The relationship between the line luminosity, $L_l$, and the
continuum luminosity, $L_c$, can be described by
\begin{equation}
  L_l \propto L_c^{\alpha} \ .
\label{eq:baldwin}
\end{equation}
\citet*{kinney90} find that $\alpha \approx 0.83$ for C$IV$, and
$\alpha \approx 0.88$ for Ly$\alpha$.

\begin{figure}
\begin{center}
\includegraphics[width=8cm]{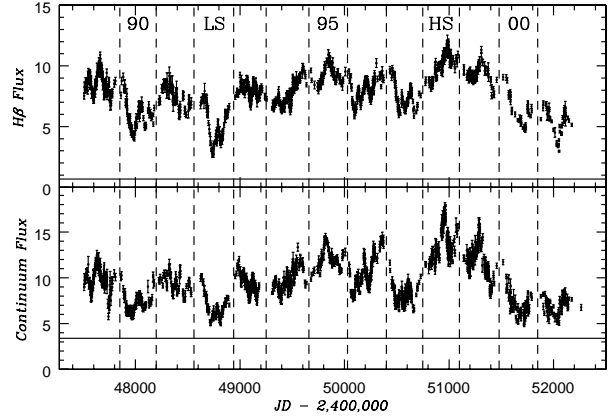}
\caption{
Lightcurves for optical continuum flux ($10^{-15}$ erg s$^{-1}$ cm$^{-2}
\rm{\AA}^{-1}$, at 5100 $\rm{\AA}$) and H$\beta$ emission line
flux ($10^{-13}$ erg s$^{-1}$ cm$^{-2}$) from the 1989-2001
\textit{AGN~Watch} data on NGC~5548.
LS and HS marks the lowest (1992) and highest (1998) states of the lightcurves
respectively.  Solid lines mark the galaxy contribution to the continuum flux
\citep{romanishin95} and the narrow-line contribution to the H$\beta$ flux
\citep{gilbertp03}.  Dashed lines separate the observing seasons.
}
\label{fig:5548}
\end{center}
\end{figure}

Within a single source, various studies have shown that
emission lines have a nonlinear response to continuum variations
\citep[e.g.][]{poggepeterson92,dietrichk95}.  This
`intrinsic Baldwin effect' \citep{kinney90,kroliketal91,poggepeterson92,
koristagoad04} where the H$\beta$ emission-line response to variations in the
continuum decreases with increasing continuum level has been observed for
NGC~5548 \citep*{gilbertp03,goadkk04}.  In terms of the continuum flux, $F_c$,
and the line flux, $F_l$, the intrinsic Baldwin effect is described by
\begin{equation}
  F_l \propto F_c^{\alpha} \ .
\label{eq:intrinbaldwin}
\end{equation}
This nonlinearity can be seen by simply examining the 13-year lightcurves
(Fig.~\ref{fig:5548}).  In the lowest state (1992), the trough in the H$\beta$
lightcurve is deeper than in the continuum lightcurve, whereas in the highest
state (1998), the H$\beta$ peak is not as pronounced as the continuum.
This argument neglects that light travel time delay smears out the emission-line
response, but this should happen to both the peak and the trough.

\begin{figure*}
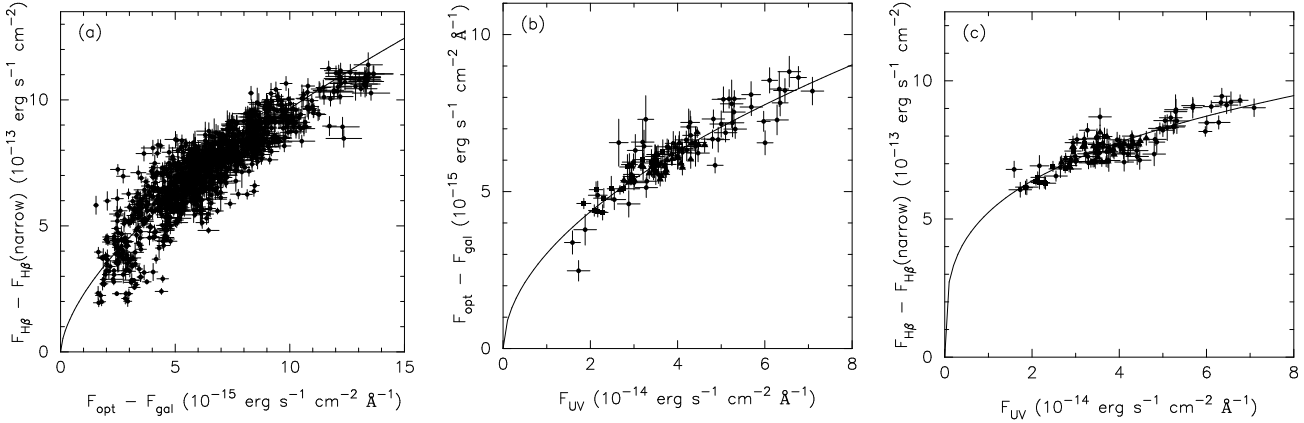

\begin{center}
  \begin{tabular}{ccc}
    \includegraphics[angle=-90,width=5.45cm]{fig2a.ps} &
    \includegraphics[angle=-90,width=5.45cm]{fig2b.ps} &
    \includegraphics[angle=-90,width=5.45cm]{fig2c.ps} \\
  \end{tabular}
\caption{
(a) Narrow-line subtracted H$\beta$ flux vs. optical continuum flux at
5100$\mathrm{\AA}$ ($F_{opt}$),
with the host galaxy continuum flux removed ($F_{gal}$ ) for the 13-year data
set. The solid line indicates
$F_{H\beta} \propto F_{opt}^{0.63}$.  A time delay between the
continuum and H$\beta$ fluxes of 17.5 days is included.
(b) UV continuum flux at 1350$\mathrm{\AA}$ vs. $F_{opt}$ - $F_{gal}$.  UV data
are taken from the IUE campaigns in 1989 \citep[circles;][]{clavel91}
and 1993 (squares) as well as
the HST campaign in 1993 \citep[triangles;][]{korista95}.  The solid line
indicates the best fit of $F_{opt} \propto F_{UV}^{0.53\pm0.02}$.
(c) Narrow-line subtracted H$\beta$ flux vs. UV continuum flux at
1350$\mathrm{\AA}$ (symbols as in (b)).  A time delay
between the H$\beta$ and UV fluxes has been added with the H$\beta$ fluxes from
1989 19.7 days behind the UV and for the 1993 data the delay is 13.6 days.
The solid line indicates the
best fit of $F_{H\beta} \propto F_{UV}^{0.28\pm0.01}$.
}
\label{fig:flux}
\end{center}
\end{figure*}

The relation between the optical continuum flux at 5100$\mathrm{\AA}$ 
and the H$\beta$ line flux is examined in
Fig.~\ref{fig:flux} (a), where a power-law (Eq.~\ref{eq:intrinbaldwin}) with
$\alpha = 0.63$ gives a good fit.
Here we corrected the optical continuum flux at 5100 $\mathrm{\AA}$, $F_{opt}$,
for the background host galaxy contribution, $F_{gal}$, where we take
$F_{gal} = 3.4 \times 10^{-15}$ erg s$^{-1}$ cm$^{-2}$ $\mathrm{\AA}^{-1}$ as
determined by \citet{romanishin95}.  The narrow-line component of the H$\beta$
line has also been removed, and we use $F_{H\beta}(narrow) = 6.7 \times
10^{-14}$ erg s$^{-1}$ cm$^{-2}$ determined by \citet{gilbertp03}.  We included
a time delay of 17.5 days
between the continuum flux and H$\beta$ flux to remove reverberation effects
which was determined by cross-correlation of the full 13-year lightcurves.
This time delay was subtracted from the times of each of the H$\beta$ data
points and the continuum flux at this new time determined via linear
interpolation.  \citet{gilbertp03} do a more detailed analysis allowing for the
different time delays observed each year, and when adopting the same galaxy
continuum background and H$\beta$ narrow-line component, determine
$\alpha = 0.65 \pm 0.02$. 
\citet{goadkk04} find that the slope of this relation is not constant,
but decreases as the continuum flux increases, an effect which is predicted by
the photoionization models of \citet{koristagoad04}.
However, the driving ionizing continuum maybe closer to that observed in the UV
at 1350 $\mathrm{\AA}$, and so previous observations of NGC~5548 at this
wavelength by the \textit{AGN Watch} using IUE and HST
\citep{clavel91,korista95} can be used to correct the relationships determined
by the optical continuum.  Using the IUE data \citet{peterson02} finds a
relation $F_{opt} \propto F_{UV}^{0.56}$, while
\citet{gilbertp03} find a relation of $F_{opt} \propto F_{UV}^{0.67}$. 
Combining both IUE and HST data, we find a relation
$F_{opt} \propto F_{UV}^{0.53 \pm 0.02}$ (see Fig.~\ref{fig:flux} (b)) assuming
no time delay between the optical and UV continuum and linearly interpolating
to get the optical continuum at the required times.  From
this result, we get a relationship between the H$\beta$ flux and the ionizing
UV flux of $F_{H\beta} \propto F_{UV}^{0.31}$.  However, the UV flux can be
combined directly with the $H\beta$ flux to determine this relationship.
Including cross-correlation time delays for the relavent years (19.7 days for
the 1989 data and 13.6 days for the 1993 data), we determine this relation
directly to be $F_{H\beta} \propto F_{UV}^{0.26 \pm 0.01}$ (see
Fig.~\ref{fig:flux} (c)).
%%%%%%%%%%%%%%%%%%%%%%%%%%%%%%%%%%%%
%%%%%%%%%%%%%%%%%%%%%%%%%%%%%%%%%%%%
%%%%%%%%%%%%%%%%%%%%%%%%%%%%%%%%%%%%
%%%%%%%%%%%%%%%%%%%%%%%%%%%%%%%%%%%%

\subsection{Luminosity dependence of time delay}\label{sec:timedep}

As the ionizing luminosity varies we expect the size of the photoionized region to
expand and contract - a larger luminosity should ionize gas to a greater
distance.
We now consider a couple of simple theoretical predictions for this effect.
If the BLR acts as a simple Str\"{o}mgren sphere with uniform gas density, then
one would predict that $R \propto L^{1/3}$.  If instead we assume that the
response in an emission line will be greatest at some density, $n$, and
ionization parameter, $\Gamma \propto L/R^2n$, then it is easy to show this
predicts $R \propto L^{1/2}$ (for a particular value of the product $\Gamma n$)
\citep{peterson02}.  More detailed photoionization modeling (using the LOC
model) by
\citet{koristagoad04} predicts a responsivity-weighted radius scaling as
$R \propto L^{0.23}$ for H$\beta$.
Photoionization models by \citet{obriengg95} and also
\citet{koristagoad04} both come to the conclusion that a relationship between
emission-line lag and incident continuum level is due to a non-linear
emission-line response.

\citet{petersonetal99} and more recently \citet{peterson02} used a
year by year cross-correlation analysis of the 
\textit{AGN Watch} data for NGC 5548 to show that the H$\beta$ emission-line
lag (relative to the optical continuum) is
correlated with the mean optical continuum flux (at 5100$\rm{\AA}$).  As the
mean optical continuum flux increases, the lag, and hence the size of the
H$\beta$ emitting region, is seen to increase. 
Using the full 13-year lightcurves for NGC 5548, \citet{peterson02} find
$\tau \propto F_{opt}^{0.95}$, though with much scatter.  They argue,
however, that the UV continuum (at 1350$\rm{\AA}$) is much closer to the
driving ionizing continuum than the optical continuum used.  Correcting for the
relationship between the optical and UV continuum (using $F_{opt} \propto
F_{UV}^{0.53}$) leads to $\tau \propto F_{UV}^{0.50}$ as predicted assuming that
the emission will be greatest at some particular gas density and ionization
parameter.

To test the predictions, using more complex methods than the
cross-correlation, we have fitted the data allowing for these breathing effects.
Firstly, we present our parameterized models and then the \texttt{MEMECHO} fits
to the 13-year (1989-2001) \textit{AGN Watch} optical continuum and H$\beta$
lightcurves for NGC~5548 before discussing these results, their findings and
implications.

\section{Parameterized models}\label{sec:param}

In this method we model the delay map and include parameters to allow it to be
luminosity dependent.  We choose to model the delay map as a Gaussian in
$\ln \tau$,
\begin{equation}
  \Psi\left(\tau\right) = \frac{\Psi_0}{A} \exp \left(-\frac{1}{2}
  \left[\frac{\ln(\tau/\tau_0)}
  {\Delta\ln\tau}\right]^2 \right)
\end{equation}
where $\tau_0$ is the peak of the Gaussian and $\Delta\ln\tau$
is the width of the Gaussian.  $\Psi_0$ scales the strength of the delay map. 
We include a normalisation factor, $1/A$, chosen so that
$\Psi_0 = \int\Psi\left(\tau\right) d\tau$, where,
\begin{eqnarray}
   A &=& \int_{0}^{\infty}\exp\left(-\frac{1}{2}
  \left[\frac{\ln(\tau/\tau_0)}
  {\Delta\ln\tau}\right]^2 \right)d\tau  \nonumber\\   
     &=& \sqrt{2\pi}\,(\Delta\ln\tau)\,\tau_0\,\exp\left[
         (\Delta\ln\tau)^2/2\right] \ .
\end{eqnarray}
We select this model as it ensures causality
($\Psi\left(\tau\right) = 0$ for $\tau < 0$) while
allowing parameters $\tau_0$ and $\Delta\ln\tau$ to control the centroid and
width of the delay distribution.

\begin{figure*}
\begin{center}
\includegraphics[angle=-90,width=13cm]{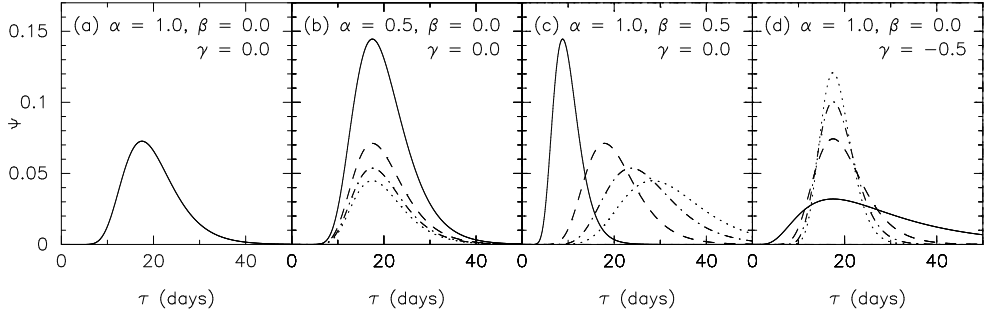}
\caption{Delay maps for different values of the `breathing' parameters $\alpha$,
$\beta$ and $\gamma$.  Different lines indicate different continuum fluxes 
where $F_{opt}$ = 5.0 (solid), 10.0 (dashed), 15.0 (dot-dashed) and 20.0 (dotted)
$\times 10^{-15}$ erg s$^{-1}$ cm$^{-2}$ $\mathrm{\AA}^{-1}$.
$\overline{\tau_0} = 17.5$ days and  $\overline{\Delta\ln\tau} = 0.3$
throughout.}
\label{fig:transfunc}
\end{center}
\end{figure*}

We introduce three `breathing' parameters to allow the delay map to be
luminosity dependent.  To account for the two effects
discussed in $\S$\ref{sec:hbdep} and $\S$\ref{sec:timedep}, we introduce
$\alpha$ to model the `intrinsic
Baldwin effect', so that $F_{H\beta} \propto F_{opt}^{\alpha}$,
and $\beta$ to allow the mean time delay of the delay map to depend on
the continuum flux i.e., $\tau \propto F_{opt}^{\beta}$.  A further parameter,
$\gamma$, is introduced to allow the width of the delay map, $\Delta\ln\tau$, to
depend on the continuum flux, so that $\Delta\ln\tau \propto F_{opt}^{\gamma}$. 
Fig.~\ref{fig:transfunc} illustrates how these parameters affect the delay map
as the continuum flux increases.  A linear line response corresponds to $\alpha
= 1.0$, whereas $\beta = 0.0$ means that there is no dependence of time-delay
with luminosity, and $\gamma = 0.0$ ensures $\Delta\ln\tau$ is constant.  A
positive value for $\beta$ indicates that the time delay is increasing with
increasing continuum flux, and similarly, a positive value for $\gamma$
indicates that the width of the delay map is increasing with increasing
continuum flux.
%%%%%%%%%%%%%%%%%%%%%%%%%%%%%%%%%%%%
%%%%%%%%%%%%%%%%%%%%%%%%%%%%%%%%%%%%

One complication is the host galaxy's contribution, $F_{gal}$, to the observed
optical continuum flux $F_{opt}(t)$.  We treat this by defining the normalised
lightcurve (with the background galaxy contribution subtracted) as
\begin{equation}
X(t) = \frac{F_{opt}(t) - F_{gal}}{\overline{F_{opt}} - F_{gal}},
\end{equation}
where $F_{gal}$ is the background galaxy continuum flux and
$\overline{F_{opt}}$ is the mean continuum flux, so that $X=1$ when
$F_{opt} = \overline{F_{opt}}$, and $X=0$ when $F_{opt} = F_{gal}$.  In all
fits we adopt a background galaxy continuum flux,
$F_{gal} = 3.4 \times 10^{-15}$
erg s$^{-1}$ cm$^{-2}$ $\mathrm{\AA}^{-1}$ as determined by
\citet{romanishin95}.

We include the three `breathing' parameters ($\alpha$, $\beta$ and $\gamma$) to
our model as follows,
\begin{eqnarray}
\Psi_0 &=& \overline{\Psi_0}~\left[X(t-\tau)\right]^{\alpha - 1} \\
\tau_0 &=& \overline{\tau_0}~\left[X(t-\tau)\right]^{\beta} \\
\Delta\ln\tau &=& \overline{\Delta\ln\tau}~\left[X(t-\tau)\right]^{\gamma} \ .
\end{eqnarray}
where $\overline{\Psi_0}$, $\overline{\tau_0}$ and $\overline{\Delta\ln\tau}$
are just the values of $\Psi_0$, $\tau_0$ and $\Delta\ln\tau$ at the mean
optical continuum flux, $\overline{F_{opt}}$.
The H$\beta$ lightcurve is then just given by
\begin{equation}
F_{H\beta}(t) = \overline{F_{H\beta}} + \int_0^{\tau_{\rm max}}
        \Psi\left[ \tau, X(t - \tau) \right]
        \Delta F_{opt}(t-\tau)
        d\tau
\ ,
\label{eq:parammod}
\end{equation}
where,
\begin{equation}
  F_{opt}(t) = \overline{F_{opt}} + \Delta F_{opt}(t)
\end{equation}

As the delay map is Gaussian in $\ln\tau$, not $\tau$, the
delay map is asymmetric and has a long tail to high time delays
(unless $\Delta\ln\tau \ll 1$). 
Although we have parameterized the delay map in terms of $\tau_0$, the lag at
which the delay map peaks, it is useful to characterise the delay map in
terms of the median lag, $\tau_{med}$, and the mean lag,
$\langle\tau\rangle$, as these values are more directly comparable to the
cross-correlation lags.  We define these quantities as
\begin{equation}
  \frac{1}{2} = \frac{\int_{0}^{\tau_{med}}\Psi\left(\tau\right) d\tau}
  {\int_{0}^{\infty}\Psi\left(\tau\right) d\tau} \ , 
\label{eq:taumed}
\end{equation}
and
\begin{equation}
  \langle \tau \rangle = \frac{\int_{0}^{\infty}\tau\Psi\left(\tau\right) d\tau}
  {\int_{0}^{\infty}\Psi\left(\tau\right) d\tau} \ .
\end{equation}

We now present tests of a series of models
allowing the delay map to vary in different ways with continuum flux. 
The models are detailed in sections \ref{sec:S} to \ref{sec:B4} with
Fig.~\ref{fig:delaymaps} showing the luminosity-dependent delay maps recovered
from these models.  The parameters found from the fits are detailed in
Table~\ref{tab:params}.  Initially, the delay map is made to be static.

\subsection{Static (S)} \label{sec:S}

The importance of allowing the delay map to be luminosity-dependent is
highlighted when fitting the data with a static delay map. In this static model
the delay map is independent of continuum level, thus we fix $\alpha = 1.0$,
$\beta = 0.0$ and $\gamma = 0.0$.  The free parameters are $\overline{\Psi_0}$,
$\overline{\tau_0}$, $\overline{\Delta\ln\tau}$ and $\overline{F_{H\beta}}$,
which are adjusted to give the best fit, determined by minimizing $\chi^2$.  The
uncertainties are derived by $\Delta \chi^2 = \chi^2_{min}/N$.  The
delay map is convolved with the continuum lightcurve to give the predicted line
lightcurve (see Eq.~\ref{eq:parammod}) with $\tau_{\rm max} = 100$ days.
However, we need to know the continuum lightcurve at all times, and
so we linearly interpolate between the continuum data points for this
purpose.  In all the parameterized model fits to the 13-year data set we take
$\overline{F_{opt}} = 9.73 \times 10^{-15}$
erg s$^{-1}$ cm$^{-2}$ $\mathrm{\AA}^{-1}$,
the mean of the optical continuum flux from this data.
Not suprisingly, this static model fits the 13-year H$\beta$ lightcurve
poorly, with $\chi^2/1248 =  9.6$ (where 1248 is the number of H$\beta$ data
points).  Table~\ref{tab:params} gives the best fitting values of the
parameters for this model and the delay map is shown in
Fig.~\ref{fig:delaymaps}.

\begin{figure*}
\begin{center}
\includegraphics[angle=-90,width=16cm]{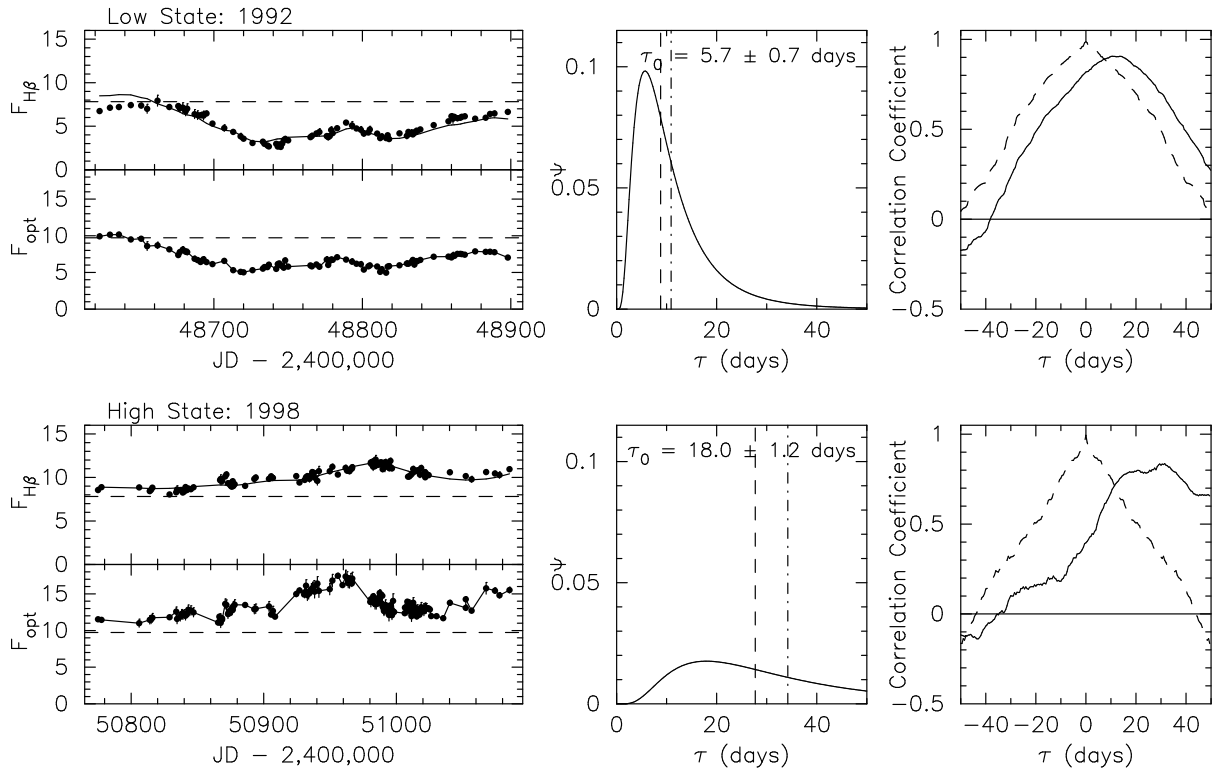}
\caption{Static delay map fit to the low state (1992, upper panel) and the high
state (1998, lower panel).  The central plots show the delay maps, where the
median and mean time delays are marked by dashed and dot-dashed lines
respectively.  The mean time delay of the delay map
in the high state is clearly larger, and the total response (the area under the
curve) is lower, than in the low state. In these fits
$\overline{\Delta\ln\tau} = 0.66$.  On the lightcurves, the dashed lines
indicate the mean continuum and H$\beta$
fluxes for the 13-year data set.  Also shown (right) is the
cross-correlation function (solid line) and continuum auto-correlation function
(dashed line) for these years.}
\label{fig:lowhigh}
\end{center}
\end{figure*}

We examine how the delay map changes with continuum flux by fitting
this static model to the data on a year by year basis
(see Table~\ref{tab:yearly}).  When fitting the static
model to the separate years, it was found that there was often not enough
information to clearly determine $\overline{\Delta\ln\tau}$, therefore we fix
$\overline{\Delta\ln\tau} = 0.66$, the value that is determined from the global
fit to the
13-year data set.  Fig.~\ref{fig:lowhigh} shows the change in the delay map
between the lowest state ($\overline{F_{opt}} = 6.7 \times 10^{-15}$
 erg s$^{-1}$ cm$^{-2}$ $\mathrm{\AA}^{-1}$ in 1992) and the highest state
($\overline{F_{opt}} = 13.5 \times 10^{-15}$
erg s$^{-1}$ cm$^{-2}$ $\mathrm{\AA}^{-1}$
in 1998).  The delay peak increases by a factor of $\sim3$ from 5.7d to 18.0d in
time delay, while the height of the peak drops a factor of $\sim5$ from 0.1 to
0.02.   Thus, the H$\beta$ response (the area under the curve) decreases by a
factor $\sim\frac{3}{5}$. 
In this figure the cross-correlation function (CCF) and continuum
auto-correlation functions (ACF) are shown for both of these years for
comparison with the delay map.  The CCF peak is close to the median of the delay
map.  In
calculating the CCF and ACF we used the \citet{white94} implementation of the
interpolation cross-correlation method
\citep{gaskellsparke86,gaskellpeterson87}.

\begin{figure}
\begin{center}
\includegraphics[angle=-90,width=8cm]{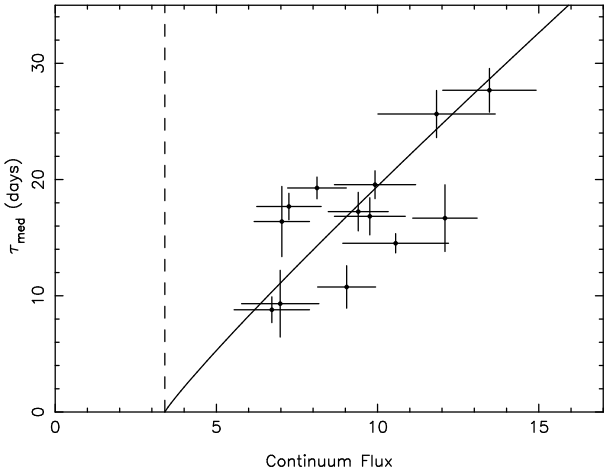}
\caption{
$\tau_{med}$ vs. mean optical continuum flux for static delay map
fit to yearly data.  $\tau_{med}$ (see Eq.~\ref{eq:taumed}) is the median of
the delay map.  Solid line shows best fitting power-law
to the data with $\tau_{med} \propto F_{opt}^{0.92 \pm 0.16}$.  Dashed line
indicates host galaxy continuum flux \citep{romanishin95}.
}
\label{fig:timedel}
\end{center}
\end{figure}

Fig.~\ref{fig:timedel} shows how the time delay increases with
increasing continuum flux.  We have plotted $\tau_{med}$ as this is less
biased by long asymmetric tail of the delay map.  Fitting a power-law of the
form $\tau_{med} \propto F_{opt}^{\beta}$ to this gives $\beta = 0.92 \pm 0.16$. 
This is similar to the \citet{peterson02} findings where the lag was
determined from cross-correlation and the resulting slope, $\beta = 0.95$ (see
their Fig. 3).  Thus, we confirm that the time delay increases with
mean continuum flux.
%%% Table of parameters for yearly static model %%%%
\begin{table*}
\begin{center}
 \caption{Parameters for fits of the Static model to the yearly H$\beta$ 
 lightcurves of NGC 5548 between 1989-2001 as well as the global fit to all
 13-years of data.  From the global fit we determine
 $\overline{\Delta\ln\tau} = 0.66$.  In the static fits we fix
 $\overline{\Delta\ln\tau}$ to this value. }
 \label{tab:yearly}
 \begin{tabular}{@{}cccccccc}
  \hline
   Year & $\overline{F_{opt}}$ & $\overline{\tau_0}$ & $ \tau_{med}$ &
   $\langle \tau \rangle$ &
   $\overline{F_{H\beta}}$ & $\overline{\Psi_0}$ & $\chi^2$/dof \\
  \hline
   1989 & $9.92\pm1.26$ & $12.7\pm0.8$ & $19.6\pm1.2$  & $24.3\pm1.5$ &
   $8.56\pm0.03$ & $0.78\pm0.08$   & $198 / 132 = 1.5$\\
   
   1990	& $7.25\pm1.00$ & $11.4\pm0.7$ & $17.7\pm1.1$ & $22.0\pm1.4$ &
    $5.69\pm0.04$ & $1.14\pm0.10$ & $278/94 = 3.0$ \\

   1991	& $9.40\pm0.93$ & $11.2\pm1.1$ & $17.2\pm1.7$ & $21.4\pm2.1$ &
   $7.22\pm0.05$ & $1.16\pm0.12$ & $88/65=1.4$ \\

   1992	 & $6.72\pm1.17$ & $5.7\pm0.7$ & $8.8\pm1.1$ & $10.9\pm1.4$ &
   $4.80\pm0.06$ & $1.15\pm0.16$ & $464/83=5.6$ \\	

   1993	 & $9.04\pm0.90$ & $7.0\pm1.2$ & $10.8\pm1.8$ &	$13.4\pm2.3$ &
   $7.78\pm0.03$ & $0.41\pm0.09$ & $386/142=2.7$ \\
   
   1994	& $9.76\pm1.10$	& $10.9\pm1.0$ & $16.8\pm1.6$ & $20.9\pm2.0$ &
   $7.68\pm0.03$ & $0.88\pm0.09$ & $324/128=2.5$ \\
   	
   1995 & $12.09\pm1.00$ & $10.8\pm1.9$ & $16.7\pm2.9$ & $20.7\pm3.6$ &
   $9.35\pm0.04$ & $0.59\pm0.11$ & $234/78=3.0$ \\
   
   1996	& $10.56\pm1.64$ & $9.4\pm0.5$ & $14.5\pm0.8$ & $18.1\pm1.0$ &
   $8.08\pm0.04$ & $0.61\pm0.09$ & $592/144=4.1$ \\
   
   1997	& $8.12\pm0.91$ & $12.5\pm0.6$ & $19.3\pm0.9$ &	$23.9\pm1.2$ &
   $7.06\pm0.04$ & $1.01\pm0.10$ & $384/95=4.0$ \\
   
   1998	& $13.47\pm1.45$ & $18.0\pm1.2$ & $27.7\pm1.9$ & $34.2\pm2.3$ &	
   $10.09\pm0.03$ & $0.65\pm0.09$ & $242/119=2.0$ \\  	
   
   1999	& $11.83\pm1.82$ & $16.6\pm1.3$ & $25.6\pm2.0$ & $31.7\pm2.5$ &
   $9.21\pm0.03$ & $0.55\pm0.08$ & $99/86=1.2$ \\
   
   2000	& $6.98\pm1.20$ & $6.0\pm1.9$ & $9.3\pm2.9$ & $11.6 \pm3.6$ &
   $5.94\pm0.07$ & $0.81\pm0.19$ & $292/37=7.9$ \\
   
   2001	&  $7.03\pm0.86$ & $10.6\pm2.0$ & $16.4\pm3.0$ & $20.4\pm3.8$ &
   $5.29\pm0.09$ & $1.31\pm0.26$ & $668/45=14.8$ \\	
   \hline
   Global & $9.73\pm2.44$ & $11.4 \pm 0.6$ & $17.6\pm1.0$ & $21.9\pm1.2$ &
   $7.67 \pm 0.02$ & $0.79 \pm 0.01$ & $12016/1248=9.6$ \\
	
  \hline
  \end{tabular}
\end{center}
\end{table*}
%%%%%%%%%%%%%%%%%%%%%%%%%%%%%%%%%%%%%%%%%%%
%%%%%%%%%%%%%%%%%%%%%%%%%%%%%%%%%%%%
%%%%%%%%%%%%%%%%%%%%%%%%%%%%%%%%%%%%
%%%%%%%%%%%%%%%%%%%%%%%%%%%%%%%%%%%%
%%%%%%%%%%%%%%%%%%%%%%%%%%%%%%%%%%%%

\subsection{B1}

In this model we allow two of the `breathing' parameters, $\alpha$ and
$\beta$ to be free, while still fixing $\gamma = 0.0$.  The delay map is now
luminosity dependent - the strength
of the H$\beta$ response and the time delay can vary with continuum flux.
The free parameters in this model are $\alpha, \beta, \overline{\Psi_0}$,
$\overline{\tau_0}$, $\overline{\Delta\ln\tau}$ and $\overline{F_{H\beta}}$.
As in the static model, we linearly interpolate the continuum lightcurve to get
the continuum flux at the required times.
This model fits the data significantly better than the
static model with $\chi^2/1248 = 6.0$ (Table~\ref{tab:params}).  The delay map
(Fig.~\ref{fig:delaymaps}) clearly shows an increase in the time delay and a
decrease in the H$\beta$ response with increasing continuum flux.
We find $\alpha = 0.66 \pm 0.03$ and $\beta = 0.28
\pm 0.05$ for this model.  It is interesting to note that $\alpha$ is lower than
the value determined by fitting the static delay map to the yearly data (Fig.
\ref{fig:timedel}).

\subsection{B2}

In an attempt to improve on model B1 we allow the width of the
Gaussian delay map to vary as a function of continuum flux, letting $\gamma$
be a free parameter in the fit.  Again, the continuum lightcurve is
linearly interpolated to get the continuum flux at the required times.
This extra parameter resulted in a slight improvement of the fit,
with $\chi^2/1248 = 5.9$.  Again, the delay map
(Fig.~\ref{fig:delaymaps}) clearly shows an increase in the time delay and a
decrease in the H$\beta$ response with increasing continuum flux.  
$\alpha = 0.57 \pm 0.02$ and $\beta = 0.41 \pm 0.08$ for this model.
$\gamma = -0.20 \pm 0.06$ allowing there to be a wider range of delays at lower
continuum flux than at higher continuum flux.

\subsection{B3}

\begin{figure*}
\begin{center}
\includegraphics[angle=-90,width=13cm]{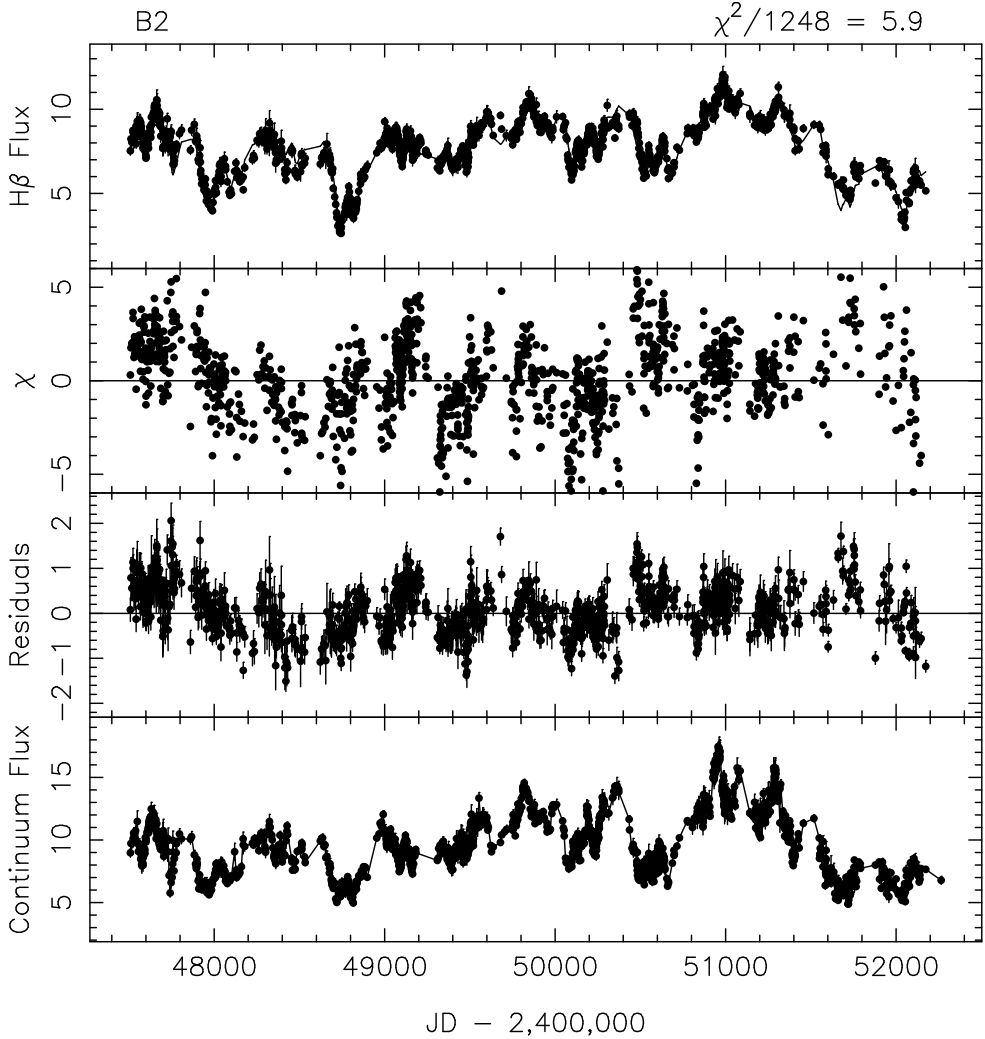}
\caption{Data and residuals for B2 model fit to the 1989-2001 H$\beta$
lightcurve.  Residuals plotted are $(data - model)$, and $\chi = (data -
model)/\sigma$.
}
\label{fig:residuals}
\end{center}
\end{figure*}

The residuals of the B2 fit (Fig.~\ref{fig:residuals}),
exhibit slow trends (timescale $\sim$1 year) that are not fit by our model.
There is no obvious correlation of these slow line variations with continuum
flux, suggesting that they are due to a process that is independent of the
reverberation effects. 
These trends may indicate a violation of the assumption that the
distribution of the line-emitting gas in the BLR is constant
over the timescale of the data.  However, as the dynamical timescale for the
H$\beta$-emitting gas, with $R \sim \tau c \sim 20$ light days
and $\Delta v \sim 5000$ km~s$^{-1}$, is $\sim 3$ years, changes in the gas
distribution may well
occur over the timespan of the observations.  To allow for this in model B3 
we fit a spline (with 26 nodes) to the residuals.  Physically, this allows the
background line flux to evolve with time e.g. accounting
for different amounts of line-emitting gas in the system.
However, \citet{goadkk04} find that the index of the intrinsic Baldwin effect, 
$\alpha$ changes on these sorts of timescales and therefore these residuals
might instead be interpreted as a consequence of that effect.
Allowing the line background flux to vary
improves the fit significantly, yielding $\chi^2/1248 = 3.1$.  The
parameters of the fit are similar to the B2 model (see Table~\ref{tab:params}),
with $\alpha = 0.58 \pm 0.03$, $\beta = 0.46 \pm 0.07$ and
$\gamma = -0.24 \pm 0.06$.

Fig.~\ref{fig:2yearly} shows the continuum and H$\beta$ lightcurves for
1991-1992 and 1998-1999 and demonstrates how well the Static, B2, and B3 models
fit the H$\beta$ lightcurve.  Particularly clear in this figure is the failure
of the static model (dot-dashed line) to fit the deepest trough and
largest peak.
%%%%%%%%%%%%%%%%%%%%%%%%%%%%%%%%%%%%
%%%%%%%%%%%%%%%%%%%%%%%%%%%%%%%%%%%%

\subsection{B4} \label{sec:B4}

\begin{figure*}
\begin{center}
\includegraphics[angle=-90,width=13cm]{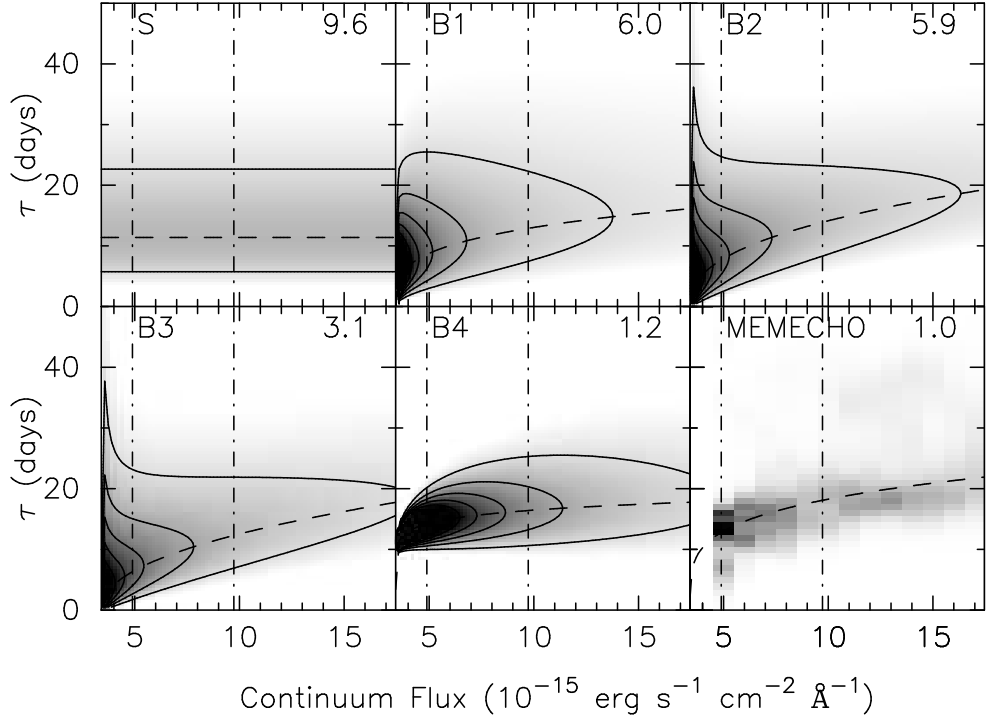}
\caption{Luminosity-dependent delay maps for the 5 parameterized models as well
as the \texttt{MEMECHO} recovered delay map.  The
reduced $\chi^2$ value of each fit is indicated. For the parameterized models,
the dashed line indicates the dependence of the peak of the
delay map on the continuum flux.  For the \texttt{MEMECHO}
model the dashed line indicates the best fit power-law, $\tau \propto
F_{opt}^{\beta}$ to the delay map.  The dot-dashed lines indicate
the minimum and mean continuum flux of the 13-year data, and the upper-limit of
the plots is the maximum continuum flux.  The lower limit of the plots is the
background continuum flux.  Solid lines mark the
contours of the delay map.  The parameters for each model are shown in
Table~\ref{tab:params}.}
\label{fig:delaymaps}
\end{center}
\end{figure*}

\begin{figure*}
\begin{center}
\includegraphics[angle=-90,width=17cm]{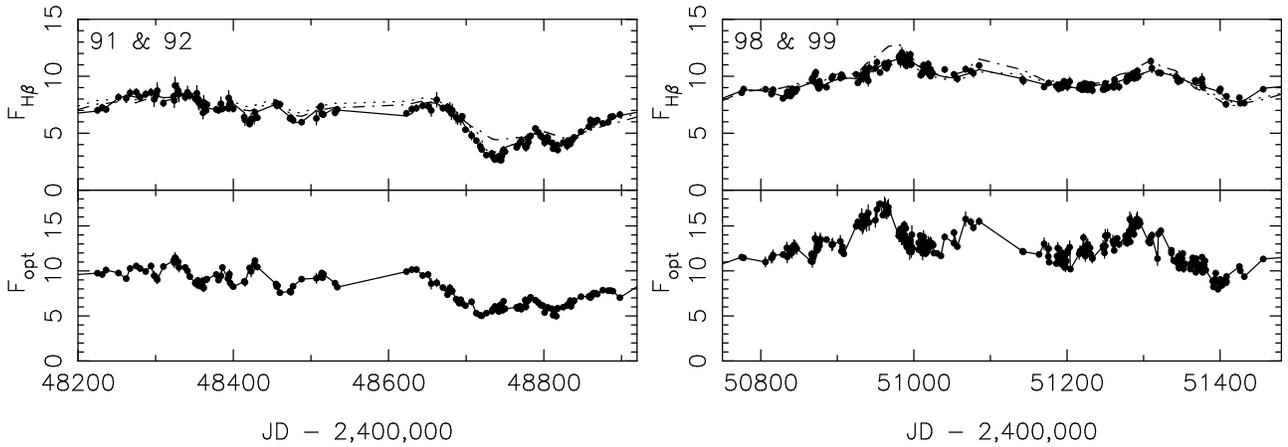}
\caption{Continuum and H$\beta$ lightcurves for 1991 - 1992 (left) and
1998 - 1999 (right).  The linearly interpolated continuum flux model is
marked with a solid line.  The fit to the H$\beta$ lightcurve can
be seen for 3 of the different models (B3 (solid line), B2 (dotted
line) and S (dot-dashed line)).}
\label{fig:2yearly}
\end{center}
\end{figure*}

In the models so far we have used linear interpolation to determine the
continuum flux at all the required times.  This, however, can lead to unphysical
lightcurves in the gaps between the data points.  It also takes no account of
the noise in the data, so a wider delay map could be being recovered in an
attempt to blur out the jagged continuum lightcurve.  The reverberation mapping
code \texttt{MEMECHO} that we also use to determine the luminosity dependent
delay map (see $\S$ \ref{sec:memecho}) uses maximum entropy methods to fit both
the continuum and H$\beta$ lightcurves.  In this model we make use of
the \texttt{MEMECHO} continuum fit to determine the continuum flux between the
data points.  The line background flux is still allowed to vary in the fit.
The extra degrees of freedom allowed in the \texttt{MEMECHO} fit
to the continuum has again improved the fit, with $\chi^2/1248 = 1.2$. 
The width of the delay map from this model is seen to be thinner
than the previous models (Fig.~\ref{fig:delaymaps}) because the
\texttt{MEMECHO} continuum lightcurve is smoother than the linearly interpolated
continuum lightcurve.  This has affected the value of $\gamma$, which is is now
positive (see Table.\ref{tab:params}), and allows the width of the delay map
to increase with increasing flux, the opposite to what was seen from the
previous model fits.
The physical interpretation of a negative value of $\gamma$ is not obvious, and
negative values of $\gamma$ found in B2 and B3 could be an artefact of using a
linearly interpolated continuum model - a wider delay map at low continuum
fluxes may be needed to smear out the jagged linearly interpolated continuum
model. However, in the B4 model the time delay still increases with increasing
continuum flux, though the relationship is flatter ($\beta = 0.10 \pm 0.01$).
$\alpha$ remains simliar to previous fits.
%%%%%%%%%%%%%%%%%%%%%%%%%%%%%%%%%%%%
%%%%%%%%%%%%%%%%%%%%%%%%%%%%%%%%%%%%
%%%% Table of parameters for all models %%%%
\begin{table*}
 \begin{center}
 \caption{Parameters for fits to 1989-2001 H$\beta$ lightcurve of NGC 5548.
 In all models the mean optical continuum level
 $\overline{F_{opt}} = 9.73 \times 10^{-15}$
 erg s$^{-1}$ cm$^{-2}$ $\mathrm{\AA}^{-1}$.}
 \label{tab:params}
 \begin{tabular}{@{}lcccccccc}
  \hline
   & $\chi^2/1248$ & $\alpha$ & $\beta$
        & $\gamma$ & $\overline{\tau_0}$ &
	$\overline{\Delta\ln\tau}$ & $\overline{F_{H\beta}}$ &
	$\overline{\Psi_0}$ \\
  \hline
  S & 9.6 & $ 1.0 $ & $ 0.0 $ & $ 0.0 $ & $11.4 \pm 0.6$
         & $0.66 \pm 0.05$ & $7.67 \pm 0.02$ & $0.79 \pm 0.01$\\
  B1 & 6.0 & $0.66 \pm 0.03$ & $0.28 \pm 0.05$ & $0.0$ 
         & $12.9 \pm 1.0$ & $0.70 \pm 0.07$ & $7.98 \pm 0.02$
	 & $0.77 \pm 0.05$\\
  B2  &5.9& $0.57\pm0.02$ & $0.41\pm0.08$ & $-0.20\pm0.06$ 
         & $13.9\pm0.8$ & $0.56\pm0.05$ & $7.98\pm0.02$ & $0.68\pm0.01$\\
  B3  & 3.1 & $0.58\pm0.03$ & $0.46\pm0.07$ & $-0.24\pm0.06$ 
         & $12.3\pm0.7$ & $0.58\pm0.04$ & $7.97\pm0.01$ & $0.68\pm0.01$\\
  B4  &1.2& $0.64\pm0.01$ & $0.10\pm0.01$ & $0.46\pm0.05$ 
         & $16.4\pm0.3$ & $0.32\pm0.02$ & $8.05\pm0.01$ & $0.68\pm0.01$\\
  \hline
  \end{tabular}
\end{center}
\end{table*}
%%%%%%%%%%%%%%%%%%%%%%%%%%%%%%%%%%%%%%%%%%%
%%%%%%%%%%%%%%%%%%%%%%%%%%%%%%%%%%%%
%%%%%%%%%%%%%%%%%%%%%%%%%%%%%%%%%%%%

\section{\texttt{MEMECHO} fit to the NGC~5548 AGN~Watch 1989-2001
lightcurves}\label{sec:memecho}

The echo mapping computer code \texttt{MEMECHO}
\citep[see][for technical details]{hornewp91,horne94} can
allow luminosity-dependent delay maps to be made
by means of a maximum-entropy deconvolution of Eq.~\ref{eq:lumdep}.
The maximum-entropy method finds the simplest positive image that fits the data,
balancing simplicity, measured by entropy, and realism, measured, in this case,
by $\chi^2$ \citep{horne94}.

\begin{figure*}
\begin{center}
\includegraphics[angle=-90,width=16cm]{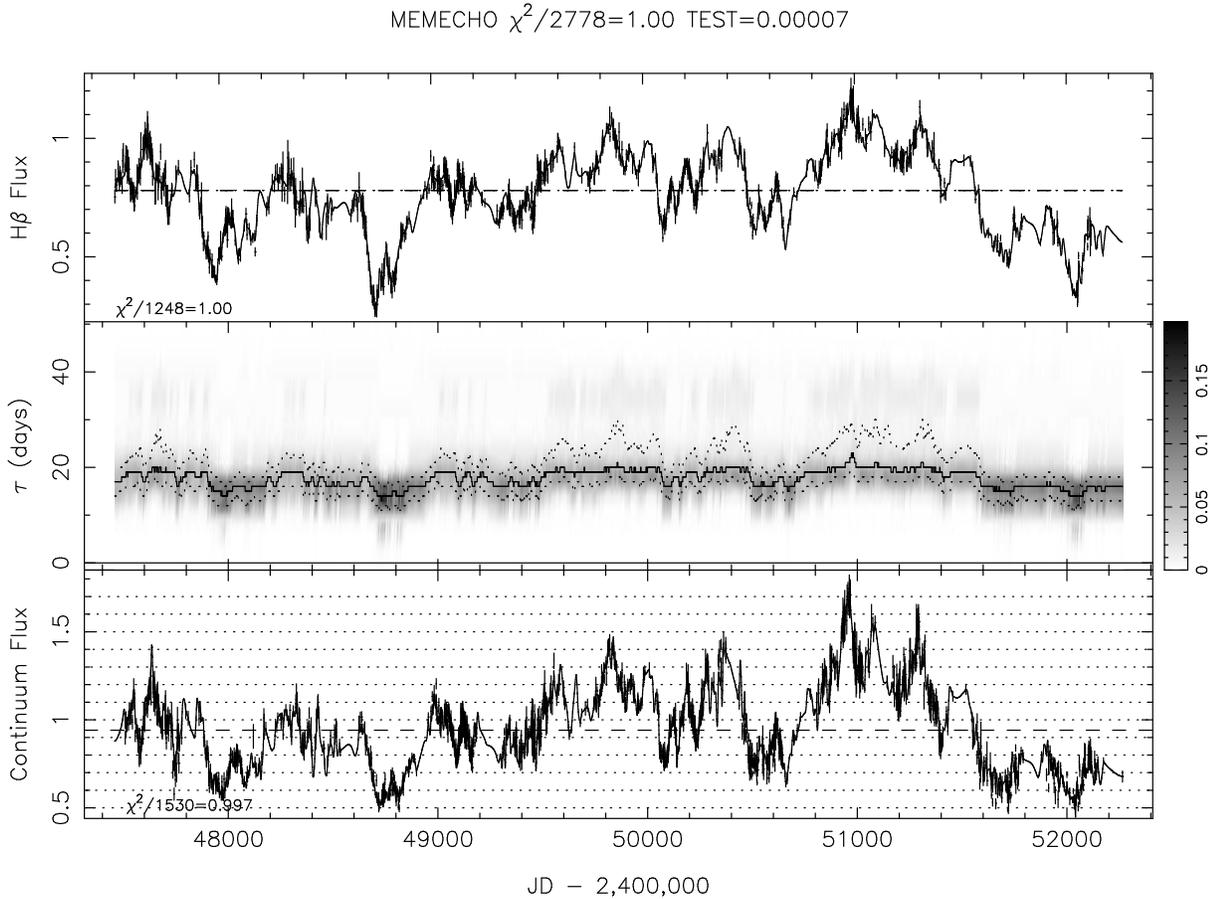}
\caption{
Lightcurves for optical continuum (bottom) and H$\beta$ emission line
flux (top) from the 1989-2001 \textit{AGN~Watch} data on NGC~5548.
The \texttt{MEMECHO} fit is the smoothest model that achieves
$\chi^2/N=1\pm\sqrt{2/N}$ for both lightcurves.  In the middle panel, the
greyscale shows the delay map at each time, the solid line marks the median of
the delay map, where as the dotted lines mark the lower and upper quartiles. 
The horizontal dotted lines in the bottom panel show the flux levels at which
each delay map is determined. 
}
\label{fig:memfit}
\end{center}
\end{figure*}

The optical continuum lightcurve, $F_{opt}(t)$, must thread through
the measured continuum fluxes, and the predicted H$\beta$ emission-line
lightcurve, $F_{H\beta}(t)$, must similarly fit the measured line fluxes.
In our fit to the data points we adjust the line background flux
$\overline{F_{H\beta}}$,
the continuum variations $F_{opt}(t)$, and the
delay map $\Psi(\tau, F_{opt})$, we take $\overline{F_{opt}}$ as the median of
the continuum data. The fit is required to have a reduced
$\chi^2/N = 1 \pm \sqrt{2/N}$, where $N$ is the number of data points.
We require this to hold for the line flux measurements,
and also for the continuum flux measurements.
The continuum is split into several flux levels (indicated in the lower panel of
Fig.~\ref{fig:memfit}), and delay maps corresponding to each level are
determined.  When computing the convolution (Eq.~\ref{eq:lumdep}) 
we linearly interpolate between these continuum levels to find the delay map
that applies to the continuum flux at time $t - \tau$.  Such an approach allows
for fully non-linear line responses.  Further details of this
can be found in \citet{horne94}.

\begin{figure}
\begin{center}
\includegraphics[angle=-90,width=8cm]{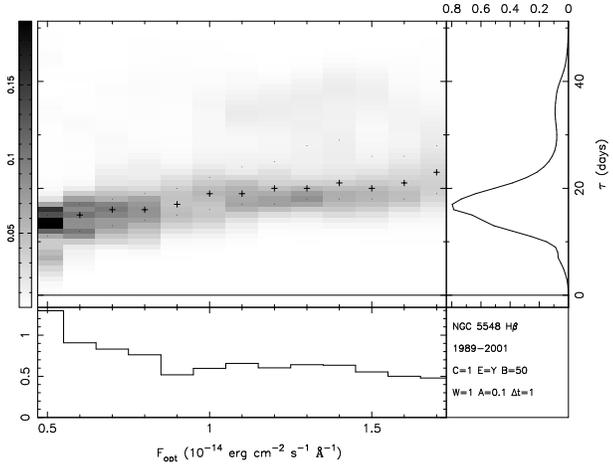}
\caption{
The luminosity-dependent delay map $\Psi(\tau,F_{opt})$ from the
\texttt{MEMECHO} fit to the 1989-2001 \textit{AGN~Watch} data on NGC~5548.
}
\label{fig:delmap}
\end{center}
\end{figure}

Fig.~\ref{fig:memfit} shows the measured lightcurves and
the \texttt{MEMECHO} fit.  The greyscale (middle panel) shows the delay map at
each time.
Fig.~\ref{fig:delmap} shows the luminosity-dependent
delay map $\Psi(\tau,F_{opt})$ (greyscale) reconstructed from
the observed lightcurves at each continuum level.  The crosses indicate the
median
of the delay map and the dots indicate the upper and
lower quartiles (see Tab.~\ref{tab:memdata}).  The lower left panel projects
the delay map along the
time delay axis and thus indicates that the amplitude of the line response
decreases with rising continuum level.  The upper right panel
gives the luminosity-averaged one-dimensional
transfer function.
At each continuum luminosity the range of delays is
relatively narrow compared to the one-dimensional transfer function.
As the luminosity rises, the mean delay increases.
At minimum light the median delay is $\sim13$~days,
and this rises to $\sim23$~days at maximum light.
Fitting a power-law of the form $\tau \propto F_{opt}^{\beta}$ (with the galaxy
background continuum removed) to the median time delay leads to $\beta = 0.24
\pm 0.08$.

\begin{figure}
\begin{center}
\includegraphics[angle=-90,width=8cm]{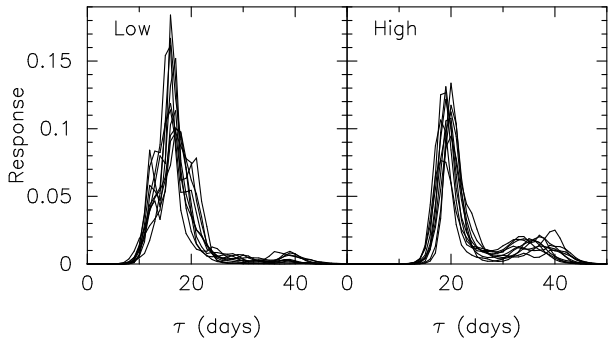}
\caption{10 realisations of the \texttt{MEMECHO} recovered delay map at low and
high continuum levels
(7.0 and $13.0 \times 10^{-15}$ erg s$^{-1}$ cm$^{-2}$ $\mathrm{\AA}^{-1}$
respectively) from Monte-Carlo simulations.
}
\label{fig:montecarlo}
\end{center}
\end{figure}

In order to understand how robust the \texttt{MEMECHO} recovered
luminosity-dependent delay map is we ran a short Monte-Carlo simulation.
We generated ten sets of continuum and H$\beta$ lightcurves with the data points
shifted by a Gaussian random number with a mean of zero and standard deviation
equal to the uncertainty on that data point.  Ten different delay maps were then
recovered using these lightcurves.  The delay maps recovered are all close to
the original.  Fig.~\ref{fig:montecarlo} shows the ten realisations of the
delay maps when the continuum flux is at the average of the low and high states
(7.0 and $13.0 \times 10^{-15}$ erg s$^{-1}$ cm$^{-2}$ $\mathrm{\AA}^{-1}$
respectively).  This shows the possible range of uncertainties expected in the
delay map. 
%%%%%%%%%%%%%%%%%%%%%%%%%%%%%%%%%%%%
%%%% Table of parameters MEMECHO %%%%
\begin{table*}
\begin{center}
 \caption{Time delay at the median, lower and upper quartile of the 
 \texttt{MEMECHO} luminosity-dependent delay map for each continuum flux.
 }
 \label{tab:memdata}
 \begin{tabular}{cccc}
  \hline
   Continuum Flux & Median delay & Lower quartile & Upper quartile \\
   (10$^{-15}$ erg s$^{-1}$ cm$^{-2}$ $\mathrm{\AA}^{-1}$) & 
   (days) & (days) & (days) \\
  \hline
  5. & 13. & 11. & 15.\\
  6. & 15. & 12. & 17.\\
  7. & 16. & 13. & 18.\\
  8. & 16. & 14. & 19.\\
  9. & 17. & 15. & 20.\\
  10. & 19. & 16. & 22.\\
  11. & 19. & 17. & 24.\\
  12. & 20. & 17. & 25.\\
  13. & 20. & 18. & 28.\\
  14. & 20. & 18. & 29.\\
  15. & 20. & 18. & 26.\\
  16. & 21. & 18. & 26.\\
  17. & 23. & 19. & 28.\\
  \hline
  \end{tabular}
  \end{center}
\end{table*}
%%%%%%%%%%%%%%%%%%%%%%%%%%%%%%%%%%%%%%%%%%%
%%%%%%%%%%%%%%%%%%%%%%%%%%%%%%%%%%%%
%%%%%%%%%%%%%%%%%%%%%%%%%%%%%%%%%%%%
%%%%%%%%%%%%%%%%%%%%%%%%%%%%%%%%%%%%
%%%%%%%%%%%%%%%%%%%%%%%%%%%%%%%%%%%%
%%%%%%%%%%%%%%%%%%%%%%%%%%%%%%%%%%%%
%%%%%%%%%%%%%%%%%%%%%%%%%%%%%%%%%%%%
%%%%%%%%%%%%%%%%%%%%%%%%%%%%%%%%%%%%

\section{Discussion}\label{sec:discuss}

Both the parameterized fit and the \texttt{MEMECHO} fit to the 13-year continuum
and H$\beta$ lightcurves for NGC 5548 show that H$\beta$ reverberations depend
upon the continuum state in such a way that greater time delays occur for
higher continuum states.
From the parameterized breathing models we find that $\tau \propto
F_{opt}^{\beta}$ with $\beta$ in the range 0.1 to 0.46, depending on the
specific model, with the continuum at 5100 $\rm{\AA}$ corrected for the galaxy
contribution.  However, as noted in $\S$\ref{sec:hbdep}, the driving
ionizing continuum may be closer to that observed at 1350 $\rm{\AA}$.  Using our
relationship that $F_{opt} \propto F_{UV}^{0.53}$, the luminosity-dependent lag
scales like $\tau \propto F_{UV}^{0.05 - 0.24}$.  The \texttt{MEMECHO} fit gives
$\tau \propto F_{opt}^{0.24}$, or $\tau \propto F_{UV}^{0.13}$ with respect to
the UV continuum.
This would appear to be in contrast with both the results of the Static model
fits to the separate years, where we find $\tau \propto F_{opt}^{0.93}$, and
the findings of \citet{peterson02}
where they find that CCF centroid lag scales as $F_{opt}^{0.95}$.
Our values from the breathing models are closer to that predicted by the
photoionization models of \citet{koristagoad04} who find that
the responsivity weighted radius, $R \propto F_c^{0.23}$, although they noted
that the power-law slope should differ somewhat in high (flatter) and low
(steeper) continuum states.

\citet{peterson02} determine the lag on a yearly basis using
cross-correlation.  As mentioned in $\S$ \ref{sec:echomap}, the CCF is the
convolution of the delay map with  the ACF.  Thus, it is clear that the time
delay between the line and continuum light curves determined via
cross-correlation depends on the shape of both the delay map and the ACF
of the continuum.  If the ACF is a delta function, then the
CCF is identical to the delay map.  Generally the ACF is broad and the delay map
is asymmetric, thus the
peak of the CCF will not necessarily occur at the same time delay as in the
delay map \citep[e.g.][]{welsh99}.  The continuum variability properties of
NGC~5548 vary from year to year leading to a change in ACF, and as an artifact
of this, a change in lag could be measured without there being a change
in the delay map.  It is also the case that the accuracy of the CCF centroid
depends on the length of the sampling window, the sampling rate, the continuum
variability characteristics and the data quality \citep{perezetal92a}. 
Throughout the 13-year monitoring campaign of NGC 5548 the mean sampling rate
has varied, which could lead to a change in the determined lag.  However, our
results from the Static fits to the yearly data produce similar lags (when
comparing our $\tau_{med}$ with the centroid lag) and a
similar slope to \citet{peterson02}, suggesting that the cross-correlation
results may be measuring the true lags.

Both the parameterized breathing fits and the \texttt{MEMECHO} fit to the full
13-year data show a much flatter dependence of the lag on the luminosity than
the year-by-year analysis shows.  
There is scatter within the results from the
yearly analysis.  For example, there is a wide range of delay ($\sim9-18$ days)
when the continuum is in a low state with
$F_{opt} \sim 7\times$10$^{-15}$ erg s$^{-1}$ cm$^{-2}$ $\mathrm{\AA}^{-1}$,
and there is also a wide range of continuum
fluxes (7 - 12$\times$10$^{-15}$ erg s$^{-1}$ cm$^{-2}$ $\mathrm{\AA}^{-1}$)
that have a lag of approximately 17 days.  A wide range of delays at low
continuum flux and also the same lag at a wide range of fluxes favours a
flatter relationship.  It could be
that these features dominate over the two years with the largest lag
when fitting to the full 13-year data set.
The tail to long delays on the delay
maps determined by the parameterized models and the \texttt{MEMECHO} fits may
also allow a flatter relationship.  The difference in slope is certainly
interesting and merits further investigation.

From both the parameterized models and the \texttt{MEMECHO} fits, we also find
that the amplitude of the H$\beta$ response
declines with increasing continuum luminosity, in other words, the change in
H$\beta$ flux relative to changes in the continuum is greater in lower
continuum states.  Detailed photoionization calculations by
\citet{koristagoad04} predict this, and \citet{gilbertp03} and
\citet{goadkk04} find the same general result.  From our parameterized fits to
the data, we find $F_{H\beta} \propto F_{opt}^{\alpha}$ with
$0.57 < \alpha < 0.66$, which is consistent with that found by
\citet{gilbertp03}.  Correcting our relationship to be relative to the UV
continuum gives $F_{H\beta} \propto F_{UV}^{0.30 - 0.35}$, which is close to the
relationship shown in Fig~\ref{fig:flux} (c).
\citet{goadkk04}  find a range in this parameter (relative to the
optical continuum) of $0.41 < \alpha < 1.0$, that correlates with the optical
continuum flux with a high degree of statistical confidence.
\citet{koristagoad04} find this to be a
natural result, given that the H$\beta$ equivalent width drops with increasing
incident photon flux.  Their model predicts slightly higher values for the
responsivity to the ionizing flux ($F_{H\beta} \propto F_{ion}^{0.54 - 0.77}$)
than we observe from the parameterized fits. 

\begin{figure*}
\begin{center}
\includegraphics[angle=-90,width=13cm]{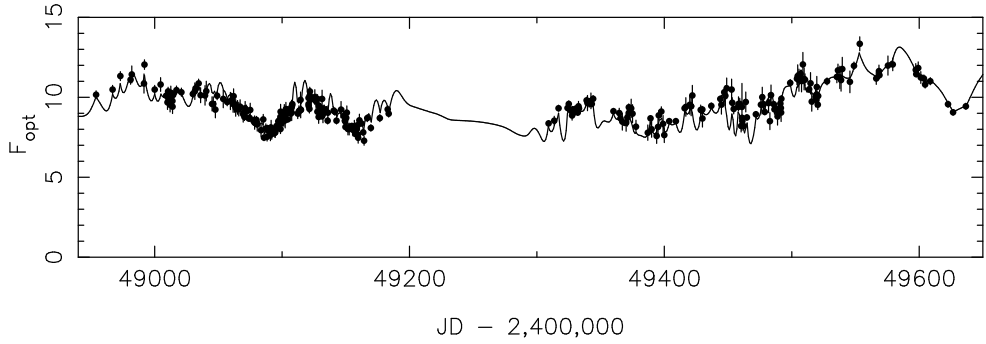}
\caption{
A section of the \texttt{MEMECHO} fit to the continuum lightcurve covering
the observing campaign in 1993 and 1994. 
}
\label{fig:memcont}
\end{center}
\end{figure*}

The \texttt{MEMECHO} recovered delay map (Fig.~\ref{fig:delaymaps}, panel (f)
and Fig.~\ref{fig:delmap}) and the delay map from the B4 model (Fig.~\ref{fig:delaymaps}, panel (e)) seem similar, which is not suprising as they are both driven by the same continuum lightcurve model.  However, the B4 model differs from the B1, B2, and B3 models, particularly in its value for $\beta$, because of the differing continuum models used. \texttt{MEMECHO} allows the continuum model to have more freedom between the data points (compared to the linear interpolation used in models B1 - B3) resulting in a delay map with a smaller width at shorter time delays as less blurring of the continuum lightcurve is required. Nevertheless, the $\alpha$ parameter does not vary too much between the different models.  It is also interesting to note that the residuals in B4 are very flat, even without the spline fit, the long term trends seen in the residuals to B2 have disappeared.  On close inspection of the \texttt{MEMECHO} fit to the continuum (see Fig.~\ref{fig:memcont}), it can be seen that between the data points the continuum model sometimes shows unphysical dips and peaks.  For instance, if the model constantly dips down between the data points, the lag between the continuum and the line lightcurves can be shifted without altering the delay map. In this way it seems that \texttt{MEMECHO} accounts for the long timescale variations that cannot be described by the delay map in the B2 model.  Although these spurious dips mean that the \texttt{MEMECHO} result may not be completely reliable, it is clear that the recovered lag luminosity relationship is much flatter than found by the cross-correlation method and thus is an important independent check of the parameterized models.

%%%%%%%%%%%%%%%%%%%%%%%%%%%%%%%%%%%%
%%%%%%%%%%%%%%%%%%%%%%%%%%%%%%%%%%%%

\begin{figure*}
\begin{center}
\includegraphics[angle=-90,width=13cm]{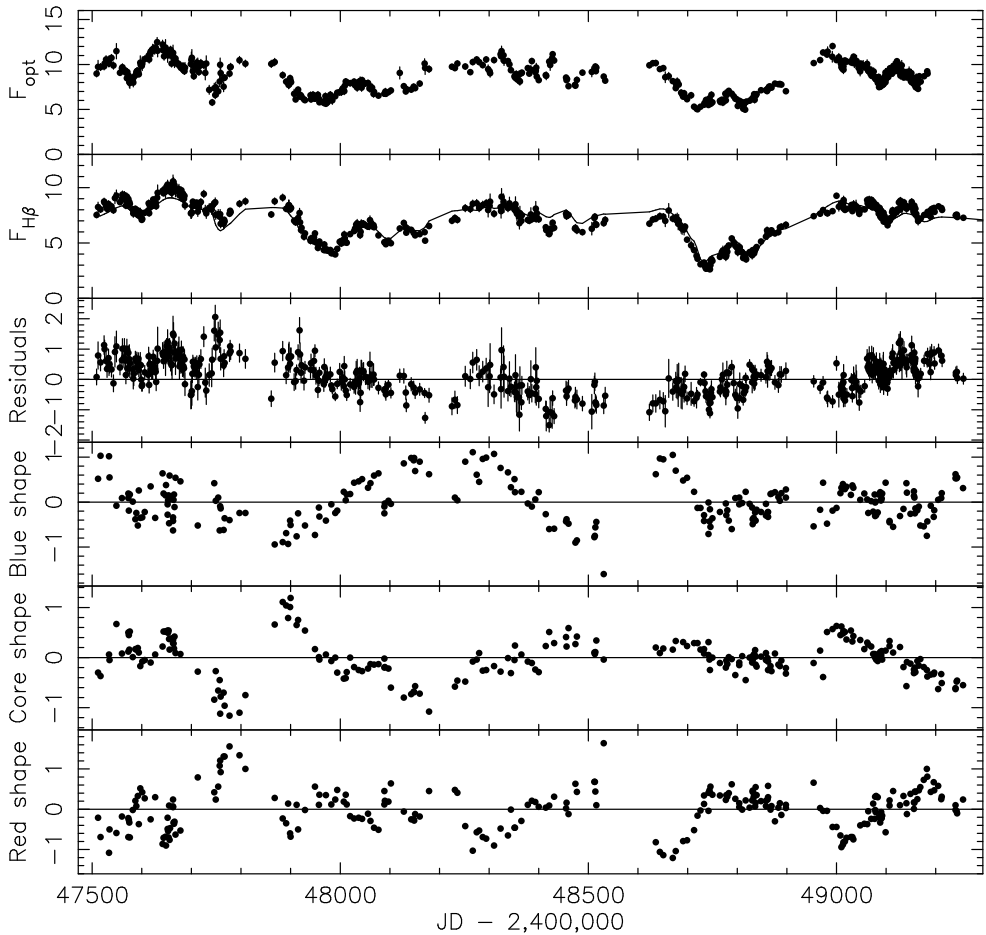}
\caption{
A comparison of the `shapes' of the red wing, blue wing and core of the
H$\beta$ line profile as calculated by \citet{wanderspeterson96} with the
residuals of the B2 parameterized model, and the continuum and H$\beta$ flux
with the B2 model shown.
}
\label{fig:shapes}
\end{center}
\end{figure*}

For the B3 and B4 models we allowed a spline fit to the residuals to
account for slowly varying trends (timescale $\sim$ 1 year, see
Fig.~\ref{fig:residuals}).  We suggest that
these long timescale variations are not due to reverberation effects, but due to
slow changes of the BLR gas distribution.  In a study of the line
profile variability of the first five years of the \textit{AGN Watch} data
on NGC 5548, \citet{wanderspeterson96} find that although the H$\beta$ emission
line flux tracks the continuum flux, the H$\beta$ emission line profile
variations do not, and therefore are not reverberation effects within the BLR.
In Fig.~\ref{fig:shapes} we compare the line profile `shapes' determined by
\citet{wanderspeterson96} for the first five years of data with our residuals
from the B2 model.  \citet{wanderspeterson96} define these emission-line
`shapes' to describe the relative prominence of features in the emission-line
profile.  Each profile is split into its red wing, core and blue wing and the
`shape' for each determined.  A positive number for the shape
indicates that it is more prominent than on average, whereas a
negative number indicates that it is less prominent than on average.
Although the shapes vary on similar timescales to the residuals, there is no
obvious correlation with the profile shape variations.  A comparison of the
profile variability for the full 13-year data with our residuals warrants
further study.
%%%%%%%%%%%%%%%%%%%%%%%%%%%%%%%%%%%%
%%%%%%%%%%%%%%%%%%%%%%%%%%%%%%%%%%%%

\section{Conclusion}\label{sec:conc}

Our analysis of 13~years of optical spectrophotometric monitoring
of the Seyfert~1 galaxy NGC~5548 from 1989 through 2001
reveals that the size of the H$\beta$ emission-line region
increases as the continuum increases.  Will also find that the strength of the
H$\beta$ response decreases with increasing continuum flux.
We have fit the H$\beta$ emission-line lightcurve using both parameterized
models and the echo-mapping code \texttt{MEMECHO}, both methods show these
effects.

In our parameterized models, we allow the delay map to be
luminosity-dependent.  Our model is parameterized such that the H$\beta$
emission can respond non-linearly to the optical continuum variations, i.e.,
$F_{H\beta} \propto F_{opt}^{\alpha}$. 
From our fits to the data we determine $0.57 < \alpha < 0.66$ which is
consistent with previous findings of \citet{gilbertp03} and \citet{goadkk04}.
However, the ionizing continuum is likely to be closer to the UV continuum
($F_{UV}$) than the optical continuum.
Correcting for this ($F_{opt} \propto F_{UV}^{0.53}$) gives
$F_{H\beta} \propto F_{UV}^{0.30 - 0.35}$.
In addition, we allow the peak of the delay map, $\tau_0$,
to be luminosity-dependent, $\tau_0 \propto F_{opt}^{\beta}$, and
find $0.10 < \beta < 0.46$.  Correcting to be with respect to
the UV continuum gives $\tau_0 \propto F_{UV}^{0.05 - 0.24}$. 
\texttt{MEMECHO} fits to the lightcurves also show these effects, and fitting a
power-law of the form $\tau \propto F_{opt}^{\beta}$ to the luminosity-dependent
delay map gives $\tau \propto F_{opt}^{0.24}$, or $\tau \propto F_{UV}^{0.13}$.
The values we determine for $\beta$ (corrected to be relative to the UV
continuum) are not consistent with the simple
Str\"{o}mgren sphere ($\beta = 1/3$) and ionization parameter ($\beta = 1/2$)
arguments and have a flatter slope than the $\beta = 0.95$ result from
the cross-correlation analysis of
\citet{peterson02}.  However, they are close
to the prediction of \citet{koristagoad04}  whose more detailed photoionization
models predict $\beta = 0.23$.

In our parameterized models we find slowly vary residuals (timescale $\sim$ 1
year) which we suggest are not reverberation effects, but indicate changes in
the gas distribution.  Comparison of these residuals with velocity profile
changes from the first five years of the data \citep{wanderspeterson96} are
inconclusive.
In future work we will examine velocity profile changes of the full 13-year
lightcurves.

\subsection*{Acknowledgements}

EMC is supported by a PPARC Studentship at the University of St Andrews.  EMC
wishes to thank Mike Goad and Kirk Korista for very useful discussions
about this work during a visit to the University of Southampton.  The authors
would also like to thank Kirk Korista, Brad Peterson and the referee, Andrew
Robinson, for comments which have improved the manuscript.
 
\bibliographystyle{mn2e}
\bibliography{iau_journals,agn}

%\bsp

\end{document}